# DESIGN AND IMPLEMENTATION OF INTELLIGENT COMMUNITY SYSTEM BASED ON THIN CLIENT AND CLOUD COMPUTING

Weitao Xu, Dongfeng Yuan and Liangfei Xue

School of Information Science and Engineering, Shandong University, Jinan, China

## ABSTRACT

*With the continuous development of science and technology, the intelligent development of community system becomes a trend. Meanwhile, smart mobile devices and cloud computing technology are increasingly used in intelligent information systems; however, smart mobile devices such as smartphone and smart pad, also known as thin clients, limited by either their capacities (CPU, memory or battery) or their network resources, do not always meet users' satisfaction in using mobile services. Mobile cloud computing, in which resource-rich virtual machines of smart mobile device are provided to a customer as a service, can be terrific solution for expanding the limitation of real smart mobile device, but the resources utilization rate is low and the information cannot be shared easily. To address the problems above, this paper proposes an information system for intelligent community, which is composed of thin clients, wide band network and cloud computing servers. On one hand, the thin clients with the characteristics of energy efficiency, high robustness and high computing capacity can efficiently avoid the problems encountered in the PC architecture and mobile devices. On the other hand, the cloud computing servers in the proposed information system solve the problems of resource sharing barriers. Finally, the system is built in real environments to evaluate the performance. We deploy the proposed system in a community with more than 2000 residents, and it is demonstrated that the proposed system is robust and efficient.*

## 1. INTRODUCTION

At the beginning of the computer era, users logged on to one centralized mainframe computer that had all the processing power, using simple terminals. Over the years there was a shift to a more decentralized computing model with powerful personal desktops. In the same period the Internet infrastructure was deployed, resulting in a highly interconnected worldwide network of servers and multimedia desktops. This evolution gave rise to a host of new problems. Viruses, Trojan horses and worms pose a daily threat to vulnerable users in their home environment. Regular software and hardware updates make up a significant part of the IT budget of both professional and home users.

Thin client computing can offer a solution for these problems. In this paradigm, all applications are executed on central servers. The client device sends events (keystrokes, mouse movements, etc.) to the server, which processes the commands, renders the appropriate graphical output and sends the images back to the client. The client only needs to decode the graphical data. This results in lighter devices, since all calculation logic can be stripped from the device. Furthermore, the cost of end-devices will decrease, which is in the interest of both private and professional users. Especially for the latter, every reduction in hardware costs, even a minor one, is a significant gain. The maintenance cost can also be reduced since managing and securing one central server farm is much simpler.

Recently, the massive growth of mobile devices has led to a significant change in the users' computer and internet usage, along with the dramatic development of mobile services, or mobile computing. Smart mobile devices such as smartphone and smart pad have rapidly pervaded our everyday life, however these devices are limited by either their capacities (CPU, memory or battery) or their network resources, meanwhile, because of their relatively high prices and small display screen, these devices are also inconvenient to be deployed in large areas such as community, offices and factories. Besides, existing cloud computing technique for this kind of service is very poor to meet the attractive quality.Therefore, developing a thin client which is energy efficient and robust is very practical. Also, the developed thin client should adaptively match with the cloud computing center and provide the services and information fast, conveniently and accurately. How to get the resource and storage according to the demand is a critical and realistic problem.

Faced with these problems, this paper proposes an intelligent community information system, which is based on the architecture of thin clients, wide band network and cloud computing servers, aiming at the direction of new information technologies represented by cloud computing, intelligent terminals, embedded system and etc. Furthermore, this paper sets up the model of user requirements and the model of integrating software and hardware in cloud computing centers.

The main contributions of this work are as follows:

1. We design and implement an information system for intelligent community composed of thin client, wide band network and cloud computing servers.

2. We implement a thin client based on ARM processor and Android System, which shows much better performance than portable devices like PDA and smartphone, in the proposed information system.

3. We design and deploy cloud computing severs consisting of virtualization management system and distributed file storage system.

4. We evaluate our system under real environments, and deploy our system in a community consisting of more than 2000 residents; it is shown that the proposed information system can meet the large amount demands of residents and show good performance. Besides, this system can be applied in other scenarios such as education information system, rural information system.

The rest of the paper is organized as follows. Section 2 demonstrates the related works. Section 3 outlines the overall architecture and describes the specifications of proposed information system for intelligent community. The technology points of thin client and cloud computing servers in the proposed information system will be introduced in Section 4 and Section 5. The system evaluation is described in Section 6. Then the paper is concluded and future work is discussed in Section 7.

## 2. RELATED WORK

Chia-Chen Kuo et al [1] explores the issues and the techniques of enabling multimedia applications for the thin client computing. Their study on the network applications over thin clients is devoted to the universal plug-in architecture which supports multimedia applications over thin clients. It is shown that the proposed architecture not only significantly enhances the multimedia capability of thin clients but also reduces the memory consumed by the clients for various applications. Thin client systems and virtualization and the available products in the market are described in the paper [2]. First of all, it is worth noting that current advances in computing and the development of pervasive applications intensify the diversity problem, giving rise to many variations in terms of performance, environments, and device characteristics. The availability of a Middleware provides authors with an integration framework for multiple and potentially diverse computing platforms. Moreover, the synergistic use of a Middleware component and Web Services turns out to be a suitable solution to integrate different software components, to easily extend, for example, the e-learning system with new features, and to improve interoperability among different systems. In [3], a wide range of well-known and widely used thin client protocols are tested in both low-motion and high-motion scenarios. It is shown that additional functionality is required to offer a satisfying multimedia experience. Honda, Y. et al [4] aim to enable the largest possible number of users to use the service platform in a secure service usage environment, assuming the provision of services for both the mass market and business users. In that paper, they demonstrate the effectiveness of applying thin client technology to the service platform they desire to achieve. In addition, they perform quantitative and qualitative evaluation to test the technical elements required for the new service platform using a prototype thin client system to measure the feasibility of the thin client technology. In [5], authors provide a reference base for the development of methodologies tailored for personal cloud computing. Besides, they also provide security architecture for personal cloud based on the security requirement analysis. Smart mobile devices such as smartphone and smart pad have rapidly pervaded our everyday life. If the limitation of their relatively poor computing resources such as processor, memory, battery can be expanded, smart mobile devices will become more powerful tool for human life. To address this problem, Nam-Uk Kim et al [6] propose the remote control architecture for real-time, close-knit interaction between real smart mobile device and virtual one. There were a number of related studies in minimizing the limitation of thin client based on the same idea, yet none have been found efficient. Pham Phuoc Hung et al. present a new method that bases its architecture on the thin-thick client collaboration. They further introduce a strategy to optimize the data distribution, especially big data in cloud computing in [7]. Choonhwa Lee et al [15] have developed a new architecture for dynamic service discovery and delivery for a full spectrum of client devices, ranging from tiny, resource-poor devices to powerful workstations.

Similar to our approach, other research efforts have been made to integrate mobile devices and cloud computing. In [8], X. Luo suggests an idea of using cloud to improve mobile device's capability. Marinelly [9] innovates Hyrax, which allows mobile devices to use cloud computing platforms.

Note that there are three main differences between our work and theirs. First of all, they realize their work on mobile phone, as discussed above, they are not suitable for intelligent community system, and instead we design and build our thin client based on ARM processor (Cortex-A9) and Android System which is very fit for intelligent community system after experiments. Besides, we set up the model of user requirements and the model of integrating software and hardware in cloud computing centers, and we deploy visualization and distributed system on Cloud Sever to support better performance. Finally, we deploy the proposed information system in a community consisting of more than 2000 residents, and it is demonstrated that the proposed system has good performance.

## 3. SYSTEM OVERVIEW

This section provides the brief overview of our proposed intelligent community information system that we design and implement. The new information system, including thin client, wind band network and cloud computing servers, could offer kinds of services, such as information services, multimedia broadcasting, online chatting, network storage and etc., enabling the community users to enjoy the services at home. Figure 1 shows the high level architecture view of the intelligent community information system.

As described earlier, the thin client adopts the ARM architecture, carrying the embedded Android system. The thin client uses the USB mouse and keyboard as the input equipment, and uses the LCD as the output display. The interactive mode is simple for the users. Simultaneously, multiple peripheral interfaces are afforded, such as USB, network, camera and audio. The applications are deployed in the information service platform, which is based on cloud computing, can avoid the computer virus and redundancy information filtering to realize the system security. The information service platform can afford the information enquiry services, multimedia broadcasting, online chatting, network storage and etc. Self-developed virtual machine management can guarantee the effective usage of resources, the efficient response of concurrent access using the load balance. Besides, this system can be applied in other scenarios such as education information system, rural information system.

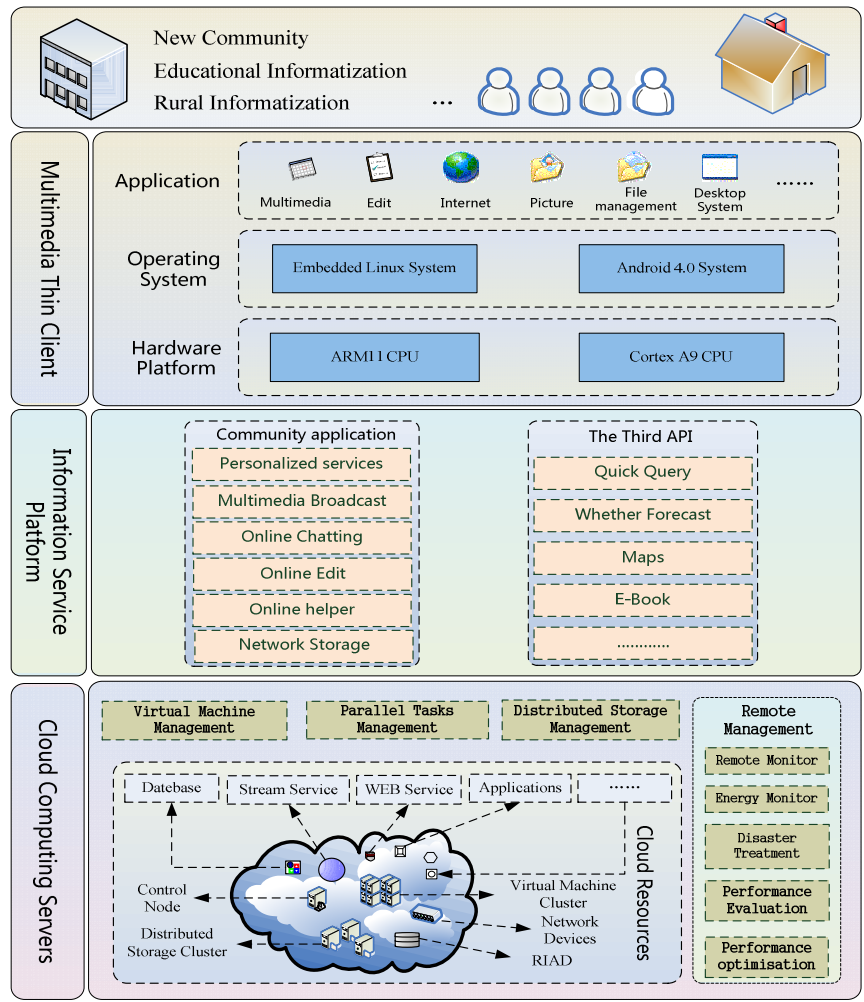

Figure 1.The overview of the intelligent community information system

## 4. DESIGN OF THIN CLIENT

Thin client systems have advanced recently with new innovative design extensions such as virtualization and cloud computing. With the development of thin client technology, desktop computers are replaced with 'thin clients' -devices that have no hard drive but instead rely on a network connection to a central cloud server where application processing and storage of information takes places [10]. Using a thin client, the user accesses a rack of blades that sits in the data center, resulting in a dedicated PC experience. With user data located in the data center, thin clients can slash break-fix costs at the client device, greatly enhance security, and eliminate the need for individual software updates and troubleshooting on desktop computers. On the server side, Virtual Desktop Infrastructure (VDI) and Server-Based Computing virtualization takes advantage of a concept long used in mainframe computing - partitioning a server so that it acts and appears as a number of independent computer devices [11, 12].

Basically, thin clients can be explained in three ways [13]. In two of the three, the architecture harks back to the early days of centralized mainframes and minicomputer. A user's machine was a terminal that processed only input and output. All data processing was performed in a centralized server.

      a) Shared Services - Input/output

Using shared terminal services software such as Windows Terminal Services and Citrix XenApp, users share the operating system and applications in the server with all other users at thin client stations. Although presented with their own desktop, users do not have the same flexibility as they do with their own PC and are limited to running prescribed applications and simple tasks such as creating folders and shortcuts [14].

      b) Desktop Virtualization - Input/output

Using products such as VMware Desktop Manager (VDM) and Citrix XenDesktop, each user's desktop (OS and applications) resides in a separate partition in the server called a "virtual machine. " Users are essentially presented with their own PC, except that it physically resides in a remote server in the datacenter. They can modify the desktop and add applications like they could with their own PC (a "Thick client").

      c) Browser-Based Applications - HTML Pages

This approach differs from the previous two in that the user's machine does the processing; however, the applications and data come from the server. The thin client contains a Web browser, and the programs come in the form of scripts on Web pages (HTML pages) from a Web server on the Internet or from the company's intranet.

In this paper, we use the third way B/S to realize thin client and applications. There are many benefits to use this approach; it makes the best use of server resources, ensuring high availability, and meanwhile, the energy cost can drop significantly. The thin client contains a web browser, and the programs come in the form of scripts on web pages from a web server on the internet. The comparison between the thin clients we designed and portable devices like smartphone, iPad will be discussed in section 7.

We have implemented a thin client based on ARM processor (Cortex-A9) and Android operating system, and this thin client typically appears to the user as simply a display unit. It is connected to a remote server for its processing power, keyboard and mouse events are sent to the server, and users can be unaware that they are using a thin client rather than a desktop PC . Figure 2 shows the hardware architecture of thin client based on ARM processor and Figure 3 presents the prototype of thin client.

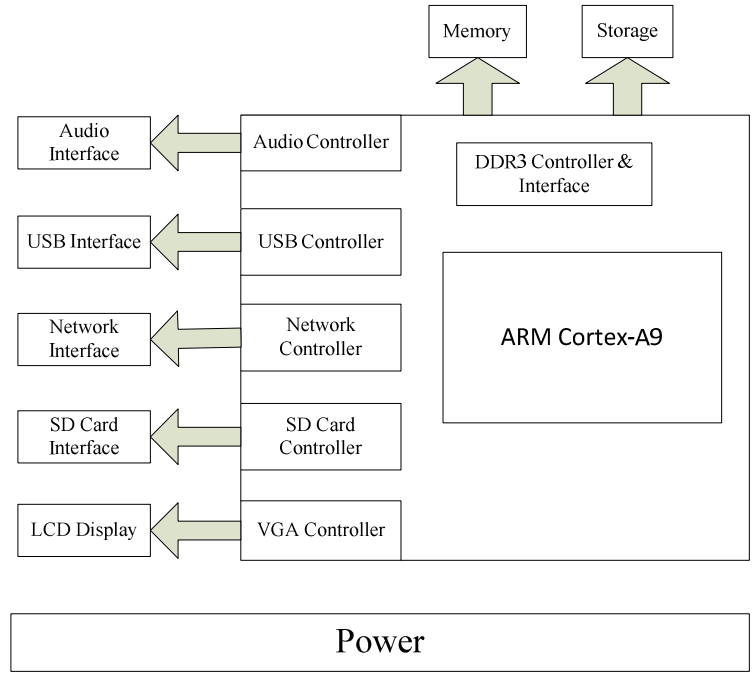

Figure 2. The hardware model of thin client

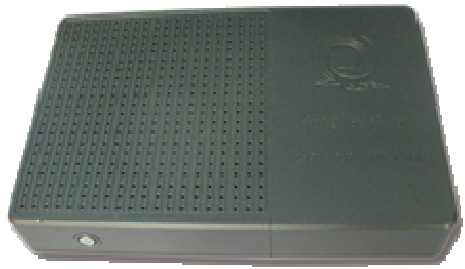

Figure 3.  The prototype of thin client

## 5. DESIGN AND DEPLOYMENT OF CLOUD COMPUTING SERVERS

The proliferation of Cloud computing has led to much industrial and academic interest, and numerous related research projects. Cloud computing services, such as Amazon's Elastic Compute Cloud, are widely available today, offering computing resources on demand. Thanks to such advances and ubiquitous network availability, the thin client computing paradigm is enjoying increasing popularity. Originally intended for wired LAN environments, this paradigm is repeating its success in a mobile context. A study from ABI Research forecasts a US$20 billion turnover surrounding services directly associated with mobile cloud computing by the end of 2014. Clearly, when applications are offloaded, the mobile terminal only needs to present audiovisual output to users and convey user input to remote servers, considerably reducing the client device's computational complexity. Consequently, applications can run as-is, without requiring (many) scaled-down versions for mobile devices.

Current DaaS deployments, such as the VMWare Virtual Desktop Infrastructure, are concentrated mainly in corporate environments. The availability of (virtual) computing resources distributed over the network lets providers offer desktop services in mobile wide area network (WAN) environments. Here, we propose virtualization management system and distributed file storage system that address the challenges providers face in offering cloud based services.

## 5.1. Virtualization Management System

The traditional virtualization management system needs high configuration of the management node in terms of the software and hardware. We develop a virtualization management system based on B/S architecture, including web server, management client, virtual machine cluster and shared storage. The architecture of the system is shown as Figure 4. Firstly, the traditional management node is replaced by the web server, which reduces the demand for resources efficiently. Secondly, the traditional management system occupies a great deal of web connections. In this new B/S architecture system, we can use the form via HTTP protocols to submit the compact data to the web server, and the web server transmits the data to the virtual machine cluster after package, which will decrease the network data transfer quantity effectively. Thirdly, to monitor the virtual machines and schedule the resources in real time, we adopt the SNMP protocols to collect the information of physical machines. Finally, we set up a two dimensional analysis model to pre-manage the free resources in the virtual machine cluster according to the habits of users.

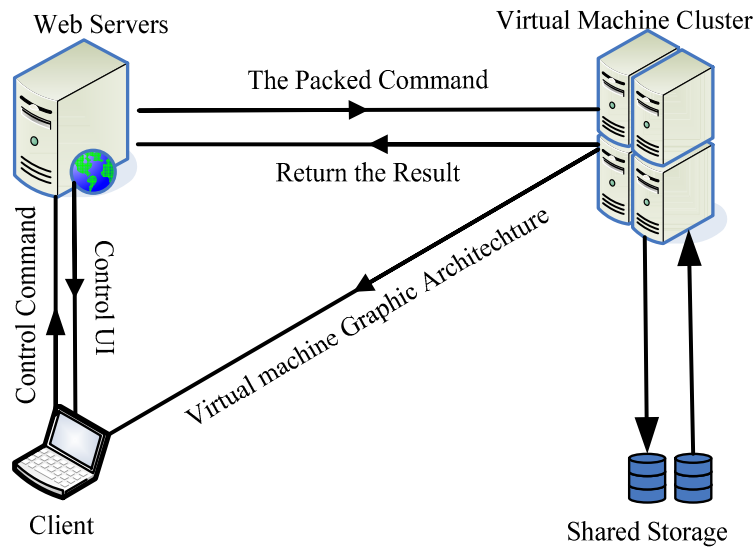

Figure 4.The virtualization management system

Compared with the similar technology products in the world, the virtualization management system based on B/S architecture can decrease the consuming of the resources and increase the efficiency effectively. Additionally, the proposed system can improve the lags of resource's scheduling.

## 5.2. Distributed File Storage System

Applications, such as streaming media and Network-drive, are necessary for the information system. The streaming media server and network file system cannot be extended dynamically, and the speed of read and write cannot satisfy the users' demand. In the new information system, the streaming media resources are stored in different nodes as many blocks. At the same time, the blocks can be backed up synchronously. When the streaming server gets the read command from the web server, the distributed file storage system will read the resources at a high speed. The streaming media distributed storage system is show in Figure 5.

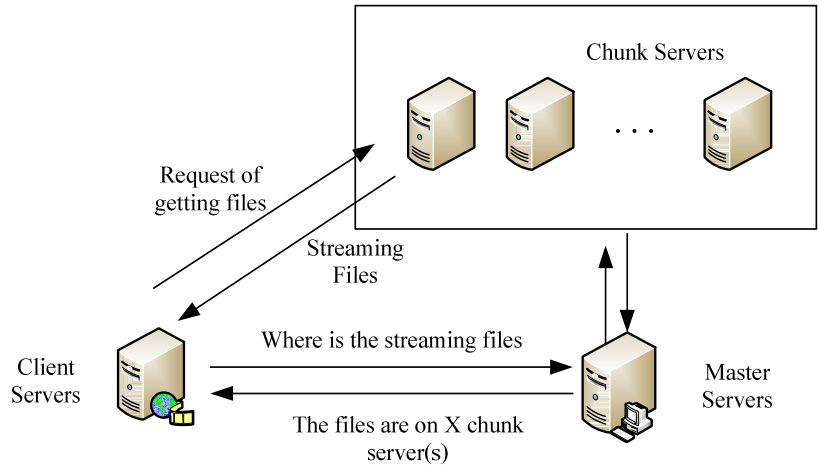

Figure 5. The Read Operation of the Distributed File Storage System

The establish of the network file system using the distributed file system can not only enhance the reliability of the data but also achieve the object of download the resources to one's own network file system from the internet directly. Compared with other systems, the files can be stored directly from the internet, which creates a seamless experience for users.

### 5.3. Focused Search Engine

The information on the Internet is too mass to mining for commonly used search engine. To solve this problem, we develop the focused search engine using an innovative fast hierarchical topic detection method. Its first step is to improve traditional topic detection by a new concept of contribution to suitable for the hierarchy. And second step is to improve traditional layered cluster algorithm based on the first step. Real result data from experiment prove that the proposed methods have better hierarchical topic detection performance and low time complexity. The web data mining model based on cloud computing can solve most users' will and be able to give reliable information service according to the request. Along with the development of cloud computing, this solution is the inevitable result in solving Web data mining. The model is shown in Figure 6.

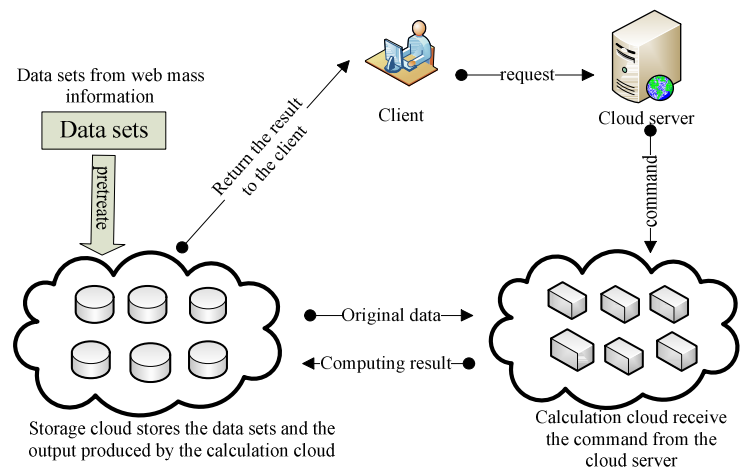

Figure 6. Web data mining cloud model

## 6. SYSTEM EVALUATION

Since the iPhone's premiere in 2007 and the follow-up booming of Android phones and tablets, we have witnessed a stunning success of mobile computing, which reflects a significant change in users' computer and internet usage. This fact leads to the presence of mobile services in almost every aspect of life, examples of which are commerce, education, health care and so on, however, these mobile devices are often relatively expensive and inconvenient to be deployed in large areas such as community, offices. Table 1 describes the comparisons between iPad, smartphone, PDA and thin client we designed for the proposed intelligent community information system, from which we can see, the computation capability of thin client is much higher than other mobile devices, and it is thinner on storage and RAM. Although thin client is not portable, it has larger display screen and can provide better user experience compared with small screen of iPad, smartphone and PDA. Most importantly, the energy cost of thin client is the lowest, which means that it's energy efficient and suitable for large scale of deployment in information system.

Table 1. Comparison between iPad, smartphone (Galaxy S3), PDA (iPAQ) and Thin Client

|  | **iPad3** | **Galaxy S3** | **HP iPAQ** | **Thin Client** |
|---|---|---|---|---|
| **CPU** | A6 1.3Ghz | Quad 1.4Ghz | PXA310 624Mhz | *Cortex-A9 1.4Ghz dual core* |
| **Storage** | 16GB | 16GB | 2GB | *2GB* |
| **RAM** | 1GB | 1GB | 64M | *1GB* |
| **Screen** | 4 inches | 4.7 inches | 3.5 inches | *19 inches LCD* |
| **Battery** | 14000 mAh | 2,100 mAh | 1,700 mAh | *1,200 mAh* |
| **OS** | iOS | Android | Windows Mobile | *Android* |

This project is supported by Shandong Province government and we deploy the proposed information system in a community with more than 2000 residents located in Jining City, Shandong Province of China, each household is given a multimedia terminal with 19 inches LCD screen, note that the cloud computing servers are located in Jinan City which is 213kilometres away from Jining City. The comparison between proposed system and traditional system is given in Table 2, as can be seen form the table, the suggested system has the characteristics of low cost, low power consuming, high reliability, high resources usage rate, easily maintenance of terminals and easily usage of the system, which solve the problem of robustness and computing resources thresholds. With the popularization of Internet and the fast development of information technology, the new system can lead to many applications in various scenarios such as education information system, rural information system.

Table 2. Comparison between Proposed System and Traditional System

| Contrast item | *The proposed information system* | PC+ traditional web access | Tablet PC/mobile phone+ traditional web access |
|---|---|---|---|
| **Terminal price** | *Low* | High | High |
| **Information** | *Low* | High | High |
| **The security of data** | *File and services are saved and backed up in the servers, which is guarded by the professionals* | Be prone to be attacked by virus. | Be prone to be attacked by virus. |
| **System maintenance** | *simple* | complex | complex |
| | *B/S, can be managed remotely* | The servers are scattered and hard to manage | The servers are scattered and hard to manage |
| **Client Power consumption** | *<=4W* | >=200W | 0.02W~10W |
| | *Cloud Computing Cluster* | High | High |
| **Resource Usage Rate** | *Above 80%* | <=20%(common users) | <=50%(common users) |
| | *Virtualization* | The resource usage rate is low | The resource usage rate is low |
| **Others** | *Be suitable for work, recreation and education* | The remote service is unavailable | Need the help of wireless |

## 7. CONCLUSIONS

In this paper, we have proposed an intelligent community information system which is composed of thin client, wide band network and cloud computing server. We design and implement thin client based on ARM processor and Android system which is fit for intelligent community information system, the designed thin client show better resources utilization than personal computer and have more powerful computing capability than mobile devices. Moreover, we design and deploy cloud computing severs consisting of virtualization management system and distributed file storage system. We apply the proposed information system in a community with more than 2000 residents located in Jining City and deploy the cloud computing servers 213Kilometers away from this community. Compared with the current PC (personal computer) or mobile devices systems, the system have a number of advantages and show good performance. Future research should be devoted to integrating relevant thin client protocol optimizations with resource allocation strategies to achieve the best user experience.

## ACKNOWLEDGEMENTS

This research was supported by Special Funding Project for Independent Innovation Achievements Transform of Shandong Province under Grant No.2009ZHZX1A0108 and No.2010ZHZX1A1001.

**AUTHORS**

Weitao Xu received his Bachelor's degree in Communication Engineering in 2006, and his Master's degree in Information System in 2010, from Shandong University. Now he is currently a Ph.D. candidate in University of Queensland. His research interests include: Sensor Networks, Embedded System, Mobile Computing.

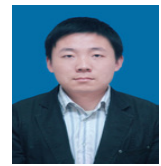

Dongfeng Yuan is currently a Professor in School of Information Science and Engineering, Shandong University, He has published over 300 papers in some technical journals and some important international conferences held by IEEE organization in his research field since 2000, his research interests include Channel modelling, Multilevel Coding (MLC) and Multistage Decoding (MSD), MIMO, Space-time coding, Turbo and LDPC codes.

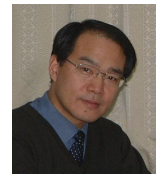

Liangfei Xue received his Bachelor's degree in Communication Engineering in 2006 from Jiangnan University, and his Master's degree in Information System in 2010, from Shandong University. His research interests include: Data Mining, Cloud Computing.

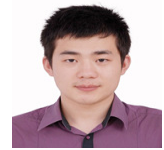